\documentclass[conference]{IEEEtran}
\IEEEoverridecommandlockouts
\usepackage{physics}
\usepackage{enumitem}
\usepackage{pgfplots}
\usetikzlibrary{patterns}
\usetikzlibrary{arrows}
\usetikzlibrary{external}
\usetikzlibrary{shapes}
\usetikzlibrary{fadings,shapes.arrows,shadows}
\usetikzlibrary{shadows.blur}
\usetikzlibrary{shapes.symbols}
\usetikzlibrary{shapes.multipart}
\usetikzlibrary{positioning}
\usepackage{subcaption}
\usepackage{qcircuit}
\usepackage{import}
\usepackage{svg}

\usepackage{cite}
\usepackage{amsmath,amssymb,amsfonts}
\usepackage{algorithmic}
\usepackage{graphicx}
\usepackage{textcomp}
\usepackage{xcolor}
\def\BibTeX{{\rm B\kern-.05em{\sc i\kern-.025em b}\kern-.08em
    T\kern-.1667em\lower.7ex\hbox{E}\kern-.125emX}}
\begin{document}

\title{Reducing Runtime Overhead via Use-Based Migration in Neutral Atom Quantum Architectures\\
\thanks{This work is funded in part by EPiQC, an NSF Expedition
in Computing, under award CCF-1730449; in part
by STAQ under award NSF Phy-1818914; in part by NSF
award 2110860; in part by the US Department of Energy Office 
of Advanced Scientific Computing Research, Accelerated 
Research for Quantum Computing Program; and in part by the 
NSF Quantum Leap Challenge Institute for Hybrid Quantum Architectures and 
Networks (NSF Award 2016136) and in part based upon work supported by the 
U.S. Department of Energy, Office of Science, National Quantum 
Information Science Research Centers. FTC is Chief Scientist for Quantum Software at ColdQuanta and an advisor to Quantum Circuits, Inc.
}
}

\author{\IEEEauthorblockN{Andrew Litteken}
\IEEEauthorblockA{\textit{Department of Computer Science} \\
\textit{University of Chicago}\\
Chicago, IL, USA \\
litteken@uchicago.edu}
\and
\IEEEauthorblockN{Jonathan M. Baker}
\IEEEauthorblockA{\textit{Department of Computer Science} \\
\textit{University of Chicago}\\
Chicago, IL, USA \\
jmbaker@uchicago.edu}
\and
\IEEEauthorblockN{Frederic T. Chong}
\IEEEauthorblockA{\textit{Department of Computer Science} \\
\textit{University of Chicago}\\
Chicago, IL, USA \\
chong@cs.uchicago.edu}
}

\maketitle

\begin{abstract}
  Neutral atoms are a promising choice for scalable quantum computing architectures. Features such as long distance interactions and native multiqubit gates offer reductions in communication costs and operation count. However, the trapped atoms used as qubits can be lost over the course of computation and due to adverse environmental factors.  The value of a lost computation qubit cannot be recovered and requires the reloading of the array and rerunning of the computation, greatly increasing the number of runs of a circuit. Software mitigation strategies exist \cite{na-comp} but exhaust the original mapped locations of the circuit slowly and create more spread out clusters of qubits across the architecture decreasing the probability of success. We increase flexibility by developing strategies that find all reachable qubits, rather only adjacent hardware qubits. Second, we divide the architecture into separate sections, and run the circuit in each section, free of lost atoms.  Provided the architecture is large enough, this resets the circuit without having to reload the entire architecture. This increases the number of effective shots before reloading by a factor of two for a circuit that utilizes 30\% of the architecture.  We also explore using these sections to parallelize execution of circuits, reducing the overall runtime by a total 50\% for 30 qubit circuit.  These techniques contribute to a dynamic new set of strategies to combat the detrimental effects of lost computational space.
\end{abstract}


\section{Introduction}

If realized physically, scalable quantum computing could dramatically affect what can be realistically computed. However, there is no obvious choice for quantum architectures as we scale from Noisy Intermediate Scale Quantum (NISQ) computing era and into fault tolerating quantum computing \cite{preskill}.  There are several technologies in different phases of development including superconducting \cite{oliver1}, trapped ion \cite{trapped-ion-challenges} and neutral atom \cite{saffman} based architectures.  All have shared challenges such as maximizing device and operation quality, but each comes with their with unique scalability challenges.  For superconducting based architectures, fabrication consistency is a limiting factor \cite{oliver1}.  Trapped ions face a similar issue when connecting different ``chains'' of qubits \cite{trapped-ion-challenges}.  Larger neutral atom architectures can lose atoms over the course of computation.  These are aspects of physical quantum computation that must be resolved to realize scalable architectures.

\begin{figure}
    \centering
    \scalebox{0.75}{
        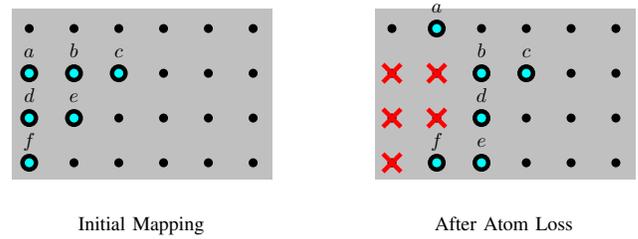%
    }
    \caption{With current methods, as circuits are adapted to lost atoms, qubits are spread across the architecture without high usage of the architecture. This adds communication, reduces probability of success and prevents full utilization of the architecture prior to reloading the array.}
    \label{fig:old-method-example}
    \vspace{-1.0em}
\end{figure}

\begin{figure}
  \centering
  \scalebox{0.72}{
        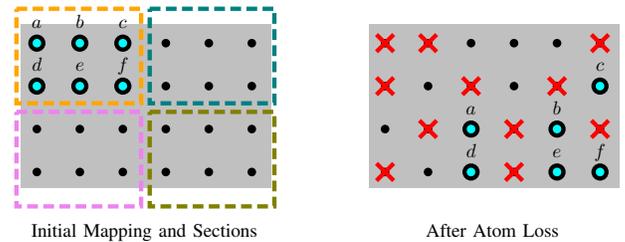%
    }
    \caption{A new relocation strategy divides the architecture into many distinct sections.  The circuit is mapped and executed in each section prior to reload. This allows for more full use of the all atoms on the architecture.}
  \label{fig:new-method-example}
\vspace{-1.5em}
\end{figure}

Neutral atom based qubits are a potential basis for scalable quantum devices.  Architectures constructed from individual atoms representing qubits, positioned in 1, 2 or 3 dimensional arrays via trapping lasers, have been demonstrated. Neutral Atom arrays in the 100s of qubits in both single \cite{Scholl_2021, Ohl_de_Mello_2019} and mixed species arrays of qubits \cite{singh2021dualelement} are in development.  Quantum circuits executed on Neutral Atom architectures are able to make use of several unique advantages including interactions with non-adjacent atoms.  As explored in \cite{na-comp}, these features can be leveraged to significantly improve circuit execution by reducing communication gate overheads and execution time.

The technology used to trap the qubits in the array is not always able to hold the atoms in place from run to run due to measurement \cite{meas-loss-1} or from outside environmental factors \cite{vacuum-atom-loss}.  This occurrence is referred to as atom loss, requiring a time-intensive reloading and retrapping of the atom array.  The extended reload time can be partially circumvented through the use of recompilation or minor reconfiguration of the compiled circuit.  In particular, minor reconfigurations can achieve similar probabilities of success when compared to full recompilation with lower overhead time.

While effective at mitigating the initial set of lost atoms to reduce overall run time, current techniques often choose to reload the array prior to exhausting all the atoms on a device due to decreases in success rate.  This is called a preventative reload.  The strategies introduced in \cite{na-comp} move the qubits away from the initial configuration of the compiled circuit introducing communication and serialization requiring preventative reloads of the array.  Figure \ref{fig:old-method-example} demonstrates how current techniques fail to make use of all the atoms on a given architecture.  If preventative reloads can be avoided, the overhead time of running a quantum circuit on a neutral atom architecture can be further reduced.

This work proposes extensions on previous software based atom loss mitigation techniques by more fully utilizing the atom array prior to a reload.  These new strategies provide significant improvements over previous methods without significantly adding to the overhead time of implementing a recovery strategy. In particular, major contributions of this works are:

\begin{itemize}[leftmargin=*]
\item Expanding on current techniques through a more flexible interaction graph which can sustain more atom loss.
\item Introducing a more deliberate approach to atom loss recovery.  We divide the architecture into multiple sections, and perform a pseudo-reload when any target metric for reloading is met, see Figure \ref{fig:new-method-example}.  The compiled circuit is remapped into a new section before the process is repeated again. This full utilization reduces overhead time by at least 50\% for a 30 qubit circuit on a 100 qubit device when compared to previous methods and as much as 80\% for a 10 qubit circuit on the same device.
\item Exploring full parallelism and partial parallelism use by running multiple instances of a circuit in several non-overlapping sections of the architecture.  By testing several different levels of parallelism we achieve up to an additional 35\% reduction in overhead time.
\end{itemize}
\section{Background}

\subsection{Quantum Computation}
A qubit is the fundamental unit of computation in a quantum system.  In contrast to a classical system, where bits are in a 0 or an 1 value, a qubit exists in a linear superposition between the $\ket{0}$ and $\ket{1}$ basis vectors described as $\ket{\psi} = \alpha\ket{0} + \beta\ket{1}$, where $|\alpha|^2 + |\beta|^2 = 1$.  When a qubit is measured, the superposition collapses into either the classical state $\ket{0}$ with probability $|\alpha|^2$ or $\ket{1}$ with probability $|\beta|^2$.  A quantum program, usually specified as a quantum circuit, is a sequence operations on sets of qubits. The operations on a quantum circuit are gates, similar to a classical circuit. The circuits ends with a final set of measurements on each qubit resulting in a classical bitstring of the same length. As the number of qubits in the quantum program increases, the number of potential output bitstrings increases exponentially. Superposition and entanglement allow us to explore the computational space in a fundamentally different way than classical computing, providing a potential way to solve classically hard problems. A more complete introduction of quantum computing can be found at \cite{mikeike}.



\subsection{Execution}

Current devices have limited connectivity, relatively high gate error rates and low coherence times. Limited connectivity requires qubit positions to be adjusted through the use of communication gates, such as the SWAP gate. Every new gate inserted into the circuit reduces the chance of producing a correct result. Additionally, the longer the time required to execute these gates, the higher the probability that a qubit will be be unable to maintain its state.  These errors prevent a single measurement of the system from being useful.  Instead, a circuit must be executed many, potentially thousands, of times to generate a distribution of results.  One of these executions is referred to as a \textit{shot}.  These shots consist of an initialization of the qubits, an execution of the circuit, and a measurement into a classical bitstring.  The most likely values from the measurement distribution are considered the answers from a run of the quantum program.

\subsection{Compilation}
Architecturally defined constraints must be taken into account when compiling a quantum circuit for a given architecture.  First and foremost, the compiler must adapt the given circuit to be executable on the hardware's topology.  Some operations are not executable if they interact too many qubits simultaneously, but can be remedied by decomposing them into simpler, lower-arity gates \cite{decomposition}. Given a hardware-defined universal gate-set there always exists a decomposition. If gates in the input circuit are not written in this gate set, it can be rewritten into this gate-set if it is universal.

Connectivity constraints are typically satisfied by inserting SWAP gates into the circuit to shift the location of a qubit from one hardware qubit to another.  Ideally, this is done with as few gates as possible to avoid error.  The process of mapping and moving qubits on a device has been explored many times, and many successful strategies have been developed to achieve low SWAP insertion \cite{map1, map2, routing1, routing2}. These generally use heuristics to map program qubits to hardware qubits (a mapping) and route qubits to keep qubits that interact often close together. These mechanisms attempt to introduce as few sources of error as possible, usually in the form of additional gates. Accurately estimating error for a quantum system is often difficult and time intensive. Therefore, any transformations of a quantum circuit typically aim to reduce proxy metrics, such as gate count, circuit depth and runtime.


\begin{figure*}
    \centering
    \scalebox{0.6}{
        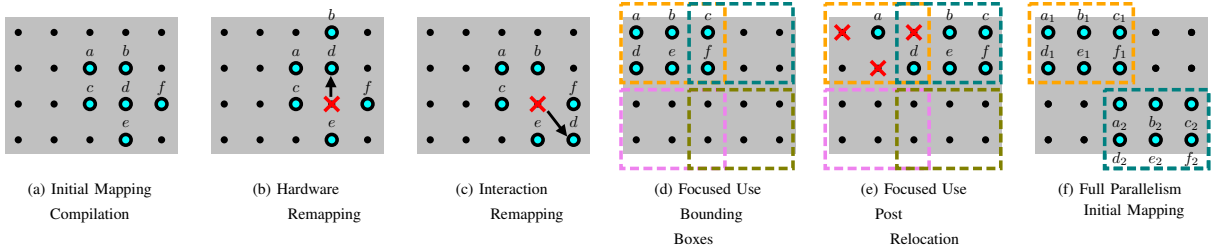%
    }
    \caption{Examples of four different atom loss coping strategies.  (a) Shows the initial configuration of qubits using the original compiler that makes use of the entirety of the architecture.  (b) shows the result of using the hardware graph to shift the qubits away from a newly lost atom.  (c) shows how the interaction graph could make a different decision, taking the shortest interaction path to an open qubit. (d) demonstrates how the architecture can be divided such that we have multiple opportunities to start in a section with fewer lost qubits.  (e) shows how a wear levelling approach would appear after one remapping of the circuit. (f) shows how these divided sections can be used to exploit some level of parallelism in the architecture.}
    \label{fig:loss-strategies}
\vspace{-1.0em}
\end{figure*}

\section{Neutral Atoms Architectures}

\subsection{Construction}
Neutral atom architectures use an atom as the computational unit to represent a qubit.  Atoms are trapped via reconfigurable optical tweezers and can be arranged in arbitrary arrays of any dimensions \cite{Graham_2019}. There have been demonstrations of arbitrary configurations \cite{Henriet}, and neutral atom arrays of up to 512 qubits have been achieved \cite{singh2021dualelement}.  However, the process of loading the qubits into the configuration can be slow, on the order of seconds.  For Neutral Atoms $\ket{0}$ and $\ket{1}$ states are represented by different energy levels and the transitions between states are facilitated by optical tuning.  For single qubit gates, we use Raman transitions.  Multiqubit gates are more complex.  The atoms are excited into a Rydberg state and are optically coupled together, forming a strong dipole interaction between the atoms.  These interactions are not limited to adjacent qubits but can interact with any qubits within a maximum distance, defined by the device.  This has significant effects on the amount of communication required.  However, long distance interactions on Neutral Atom architecture have a drawback.  All qubits within a specific radius of the qubits involved in the interaction cannot be used without interference.  This can induce a serializing effect on the circuit, but tends to be balanced out by reduced communication costs \cite{na-comp}.

\subsection{Adapting Compilation}
The principles of quantum computation do not change on a neutral atom system, but we have a different set of tools to help us build circuits for neutral atom devices.  Compilers for neutral atom systems follow the same basic algorithm for compilation.  With longer distance interactions, this provides a much larger search space for choosing the ``best'' location to move the new qubit to. Longer interaction distances lend themselves towards shorter swap paths, lower gate counts and circuit depths, all of which the compiler favors.  The compiler also attempts to avoid serialization by packing as many operations that can be performed at the same time into the same time step.

\subsection{Atom Loss}
Atoms can be lost from the array via multiple processes. The first is due to environmental factors. Stray photons or elements of an imperfect vacuum can knock an atom out of the array rendering the atom unusable as a qubit \cite{vacuum-atom-loss}.  The second is contained to atoms that undergo measurement \cite{meas-loss-1, meas-loss-2}. More frequently, when measured the atoms can be dropped from the trapping mechanism. When an atom is lost, it cannot be measured, meaning that the particular shot where the atom was lost cannot be used towards the probability distribution.  These losses are detected via a fluorescence after each shot, a process which takes on the order of milliseconds.  Following the fluorescence, if any computational atoms have been lost, an adaptive strategy must be used to recover from the atom loss.  If the strategy fails the entire array is reloaded.

\subsection{Mitigation Techniques}
When an atom required for computation is lost from the array, there are three options: 
\begin{itemize}[leftmargin=*]
    \item \textit{Reload the Array.} This is the most naive course of action is ot reload and retrap the atoms.  Reloading the array allows each qubit to be used in the next shot.  However, this is a time intensive process.
    \item \textit{Recompile to avoid lost atoms.} We inform the compiler of the missing atoms and use the same compilation pipeline as before.  This method has the most information at its disposal and can create the highest fidelity adaptations to the adjusted architecture.  However, recompilation is more time intensive than reloading the array, more akin to running a whole new circuit, so is often not the best choice.
    \item \textit{Adapt current compilation to lost atoms.} Rather than using full recompilation, adaptation strategies attempt to strike a balance between overhead time while without adding extra communication or serialization and withstanding significant atom loss.
\end{itemize}

The most effective light-weight adaptive technique has two steps.  The first is virtual remapping.  When an atom is lost, we search each adjacent direction.  We select the direction that has the the most qubits that are not lost or previously mapped to.  The mapped qubits in the selected direction are moved one atom in that direction.  These changes are recorded.  At this point, the qubits are mapped to viable atoms.  This can be seen in \ref{fig:loss-strategies}c.  However, some operations may not be able to be executed since the newly remapped qubits could be out of range of one another and requires rerouting.  When we find a set of qubits that are outside the maximum interaction distance, we find the shortest path of non-lost atoms along the possible interaction paths.  We insert SWAPs into the circuit along the path, perform the interaction and then swap back to the original positions to maintain new mapping. This method can recover any amount of atom loss, up to the number of extra qubits available on the device as long as there is an available path on device. This strategy, as explored in \cite{na-comp} is able to efficiently reduce the overall time of running many shots of a quantum circuit and maintain similar probabilities of success as full recompilation. Additionally, we can compile circuits as if the maximum interaction distance was smaller than the actual maximum interaction distance reducing the need for any rerouting  since atoms will be moved away from qubits the interact with less frequently.
\section{Motivation}

The mitigation techniques developed in \cite{na-comp} are effective at keeping overhead time low as atoms are lost.  One potential drawback is that virtual remapping can only shift the qubits in one of four directions.  If there is no room to shift in these directions, the recovery strategy will also fail. In situations where most of the atoms in the architecture are in use or lost, virtual remapping is inflexible.  There may still be atoms available for recovery.  Atoms will be lost more frequently for large circuits, requiring many more runs of the circuit.  So, any flexibility that mitigate reloading the array would be beneficial.

Additionally, a consequence of shifting partial columns of atoms is qubit spread across the architecture.  As seen in Figure \ref{fig:old-method-example}, after many atoms are lost, the atoms have been shifted far away from their original location. Based on our mapping strategies, this also means that each qubit is further away from the qubits it interacts with.  While extra communication can fix this, distribution across the architecture is not conducive to successful execution.  This also renders the originally mapped location of the atoms unusable, so communication paths are even longer.  Rather than simply letting the circuit be mapped and routed to any piece of the architecture, having focused use of a specific sections of the architecture over time may improve probabilities of success and prevent preemptive reloads.

\section{Methods}

\subsection{New Strategies}

We build on the compiler, router, and atom loss strategies developed in \cite{na-comp}, found at \cite{github-na}, to build new methods for atom loss mitigation upon previous strategies with greater flexibility and focused use on a particular piece of an array.

\begin{itemize}[leftmargin=*]
    \item \textit{Remapping via Interactions.} Previous virtual remapping strategies have been limited to remapping atoms to directly adjacent atom on the neutral atom array.  This limits movement.  However, we can create an \textit{Interaction Graph} where every atom is a node, and edges are defined between any atom within the maximum interaction distance of that atom, seen in Figure \ref{fig:loss-strategies}c.  When an atom is lost, rather than shifting the row or column of qubits, we find the shortest path along the interaction graph to an unused atom and shift any mapped qubits towards the non-mapped atom along this path.  This widens the search space of available atoms, and will ideally move fewer qubits.
    \item \textit{Focused Use and Migration.} In the original compiler, a circuit could be mapped and routed onto any qubit in the architecture.  This gives global scope, and finds good mapping and routing solutions, but can lead to less dense mapping across the architecture.  Instead, we define a bounding box big enough to hold the circuit, seen in in Figure \ref{fig:loss-strategies}d, where for a six qubit circuit, we define a 2 by 3 bounding box.  This can be done in two ways: a loose or tight configuration. The former defines both dimension by the ceiling of the square root of the number of qubits.  The latter defines one dimension by the square root, and the second by the number of qubits divided by the square root.  This box is then tiled across the architecture.  In the event that the bounding box does not neatly fit, some overlap is allowed.  The circuit is mapped and routed entirely within one of these sections. When atoms are lost, we continue using any recovery method previously defined.  When the recovery method fails, or the estimated probability of success falls below a certain threshold, rather than resetting the array, we directly remap the circuit to a new section of the architecture, accounting for any previously lost atoms. The process after one relocation is shown in Figure \ref{fig:loss-strategies}e.  Since the atoms in the new section have not been used for computation, fewer will have been lost.  We repeat this process until each section has been visited once.  Then, the entire array is reloaded before restarting the process.
    \item \textit{Parallel Executions.} The non-overlapping sections defined for \textit{Focused Use} can be treated as if each is its own architecture.  Instead of shifting the mapped circuit from one defined region to the next, we can create a single aggregate circuit \cite{multiprogramming-qaunt} that contains multiple instances of the smaller circuit.  This is similar to \cite{concurrent-mappings} where multiple variational circuits have been mapped onto a single architecture to improve their performance.  We map the circuit onto the architecture multiple times, and run multiple shots of the individual circuit in one run of the aggregate circuit.  Then any individual circuits that lost no computational atoms can be treated as a successful shot.  If an atom is lost, we can use any recovery methods previously explored until it fails.  At this point, the array is reloaded.
    \item \textit{Focused Use + Partial Parallelism.}  Rather than filling every section with qubits, we only fill some of the sections.  Then, when the recovery strategy fails, we use a different set of the predefined sections.  We are exploiting some amount of parallelism, but are giving more resources for the more focused recovery strategies to make use of.  This process continues until each section has been used at least once.  The entire array is then reloaded.
\end{itemize}

\subsection{Benchmarks}
We examine four benchmarks which can be scaled to an arbitrary number of qubits. We study how circuit-level properties like parallelism and operation density affect the performance of our recovery techniques. Higher interaction density can be difficult to route without communication and can potentially lead to more serialization. The specific circuits we study are:
\begin{itemize}[leftmargin=*]
    \item \textit{Log-Depth Generalized Toffoli (CNU).} This circuit is an N-Controlled X gate that is generally used as a piece of another circuit \cite{qubit-toffoli}. It can be performed in $log(N)$ depth, but requires many extra ``scratch'' qubits.  The controls are grouped into pairs, and target one of the scratch qubits initialized to $\ket{0}$ with a Toffoli gate.  This process is repeated with the targets until we target the final qubit.  The same set of operation is performed in reverse.  These pieces can be performed in parallel before converging on the final target operation.
    \item \textit{Cuccaro Adder.} Another component circuit, the Cuccaro Adder takes two N-bit numbers and adds them together, requiring 2N+2 qubits in a heavily serialized circuit \cite{cuccaro}.  Corresponding qubits are grouped together in blocks of operations, resulting in a carry-out bit to be used to the next block, similar to a classical ripple-carry adder.  This circuit is also low density.
    \item \textit{Quantum Approximation Optimization Algorithm (QAOA).}  QAOA is an algorithm that is used to solve certain kinds of combinatorial problems such as the Max-Cut problem \cite{qaoa}.  We use a version of this algorithm that uses a random graph with edge density of 20\% between N qubits to inform our circuit.  Each edge inserts a pair of CX gates surrounding a Z Gate.  QAOA is a somewhat dense graph, using 20\% of $N^2$ possible interactions.  While it depends on the construction of the graph, QAOA can be relatively parallel.
    \item \textit{Linear Variational Quantum Eigensolver Iteration (VQE).}  VQE attempts to find the minimum eigenvalue of a wave function encoded in parameterized rotation gates and entangled qubits \cite{vqe}.   It is then run several times while tuning the rotation gates.  We test one of these iterations. We use linear entanglement, meaning that the first qubit targets the second, the second then targets the third, and so on.  This lends itself to very low density, highly serialized circuit.
\end{itemize}

\subsection{Atom Loss Simulation}

\subsubsection{Architectural Configuration}
We mainly focus on a representative 10 by 10 neutral atom array of qubits with varying maximum interaction distances.  We focus on this style of architecture, since in the near term, two dimensions are the most practical form for a neutral atom architecture.  We use a grid, since this form of an array allows for easy reuse of optical tweezers to capture rows of qubits. Other structures would require an additional layer of complexity.  While the distances between adjacent qubits in a mesh could vary, the distance between qubits is not the limiting factor. Rather, it is how many qubits are reachable from another qubit within the maximum interaction distance. The effect of a less dense array can be approximated via the use of a smaller interaction distance. For this architecture, we create zones of restriction around each qubit with a radius equal to half the distance between the qubits.

\subsubsection{Error Rates and Probability of Success}
At present, NISQ architectures have relatively high error rates relative to what will be required for fault-tolerant quantum computing.  Neutral atom architectures are still under heavy development and have worse error rates than other NISQ architectures.  At present, we have seen fidelities of 99.6\% for one qubit gates and 96.5\% for two qubit gates \cite{multiqubitgates}.  We estimate probability of success for circuits based on these values.

However, neutral atoms do have one significant advantage in terms of error.  In the ground state, atoms are much less likely to decohere, with $T_1$ and $T_2$ times on the order of 7 seconds and 30 seconds, respectively \cite{na-lifetimes}. When in the excited state, the error from decoherence is accounted for by gate error rates.  This is a separate problem from atom loss and can occur in both the excited state, when the qubit is being operated on, and the ground state. We use a product of two items: the product of the gate successes and the probability that no qubit decoheres. We calculate the probability of decoherence in the ground state. This is estimated via the following expression: $e^{-\Delta g/T_{1,g}-\Delta g/T_{2,g}}$ where $\Delta g$ is the time the qubit is in the ground state.

\begin{figure*}
    \centering
    \makebox[0.85\columnwidth]{
    \scalebox{0.9}{
        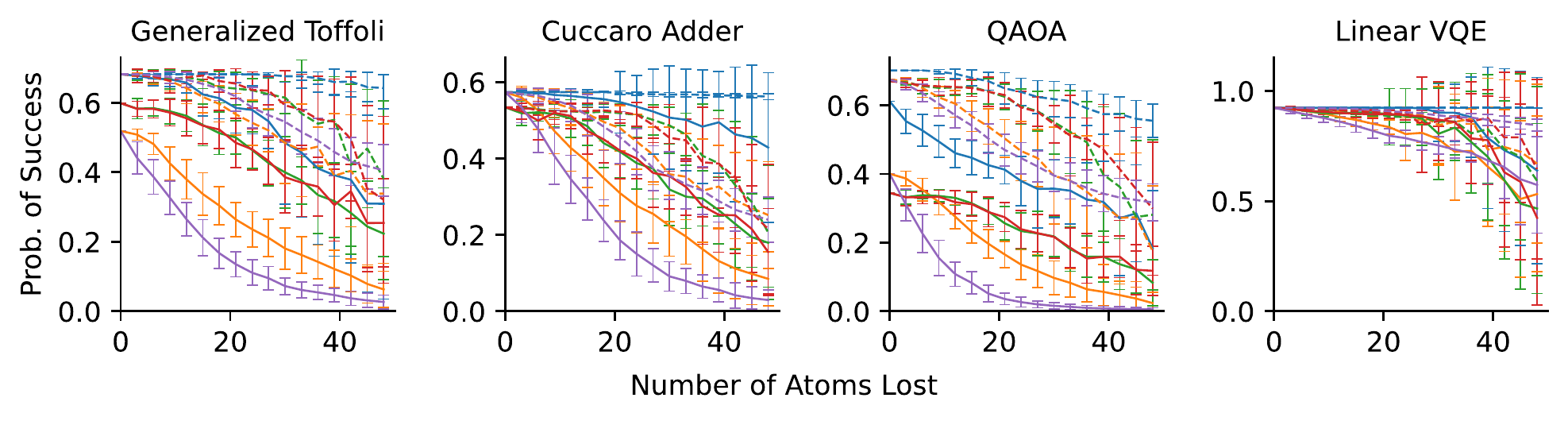%
    }}
    \vspace{-1em}
    \caption{Decreases in success rate for different mitigation strategies for 30 qubit circuits.  Each color is a different strategy, solid lines are circuits with interaction distance 3, and dash lines are interaction distance 5. Run over 50 trials, each error bar represents one standard deviation from the mean. We see decreases for all strategies, with full recompilation able to maintain the highest probability of success and relocation better approximating recompilation than our baseline strategy.}
    \label{fig:30-qubit-success-rate}
\vspace{-1.5em}
\end{figure*}

\begin{figure*}
    \centering
    \makebox[0.85\columnwidth]{
    \scalebox{0.9}{
        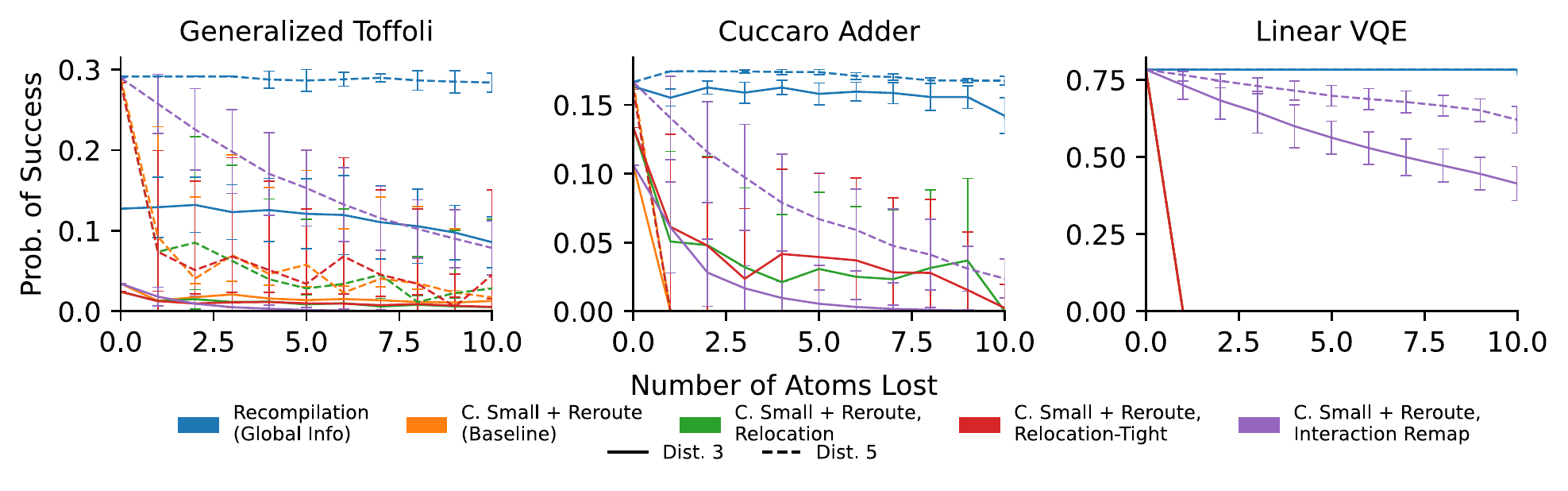%
    }}
    \caption{Decreases in success rate for different mitigation strategies for 90 qubit circuits. We exclude QAOA from this test as the success rates are not meaningful at current error rate. Each error bar represents one standard deviation from the mean over 50 trials. We see the same patterns, with the exception that interaction remapping is often able to outperform other strategies.}
    \label{fig:90-qubit-success-rate}
\vspace{-1.0em}
\end{figure*}

\subsubsection{Atom Loss Mitigation Rates}
Recall that we must consider two different kinds of atom loss: the less frequent loss due to environmental factors that applies to any atom in the array, and the more frequent loss due to interaction that only applies to atoms involved in computation.  Atom loss due to environmental factors occurs at a 0.068\% chance per atom, per shot \cite{vacuum-atom-loss}. Atom loss due to measurement occurs at a 2\% chance per atom, per shot \cite{meas-loss-2}.  After each shot, we must check if an atom has been lost through fluorescence, a process which takes 6 ms.  Reloading the entire array takes much longer, 320 ms.  This will only occur when the recovery strategy fails.  To estimate any overhead time for mitigation techniques, we consider them to be implemented in embedded hardware. Any virtual remapping is implemented in a fast, hardware lookup table which is estimated to handle reads at speeds of 40 ns, and writes as 45 ns.  Timing for mitigation algorithms are estimated by using this hardware.

All of these experiments were performed on a machine using Python 3.9 \cite{python}, Intel(R) Xeon(R) Silver 4110 2.10GHz, 132 GB of RAM, on Ubuntu 16.04 LTS.  The initial compiler and mitigation strategies were built on work from \cite{github-na}.  Any error bars show one standard deviation from the mean.
\section{Evaluation}
As we lose atoms, we want a mitigation strategy that adheres as closely to the success rate found for recompilation as possible.  Recompilation maps and routes from scratch, with access to global information to determine the best mapping and routing for the situation and best adapt to lost atoms.  We also compare to compiling to a smaller interaction distance with rerouting, one of the better strategies developed in \cite{na-comp}.

\subsection{Success Rate Resistance to Atom Loss}

The mapping and routing found for the entire architecture should have the highest probability of success and a successful mitigation strategy works to maintain this success rate.  In general, this means inserting fewer extra gates and reducing serialization induced by only moving interacting qubits further away from one another when necessary.

\subsubsection{Remapping Along Interactions}
We first examine the effects of remapping via the interaction graph rather than via directly adjacent qubits on the hardware array.  In the 30 qubit circuits, Figure \ref{fig:30-qubit-success-rate}, remapping and rerouting based only on the interaction graph fails to improve probability of success. When we remap qubits along the total hardware interaction graph, we gain flexibility, but move qubits further away from their original position.  In doing so, we must insert more communication to handle the extra distance, causing the probability of success to decrease more rapidly.

On the other hand, remapping along the hardware array often fails for larger circuits.  For the 90 qubit circuits, Figure \ref{fig:90-qubit-success-rate}, we fail to find significant average probabilities of success for several circuits after the loss of more than one or two atoms when using our original strategies.  They do not provide enough flexibility to find available atoms at this size, which is why some strategies cannot be seen on the graph. However, when we shift along the interaction graph, we are able to achieve much higher fidelities. Particularly for the Cuccaro Adder, we see that at for a maximum interaction distance of 5, the interaction model based approach achieves much higher probabilities of success. The non-recompilation counterparts are not able to sustain a minimal amount of atom loss at this size of circuit.

\begin{figure*}
    \centering
    \scalebox{0.75}{
        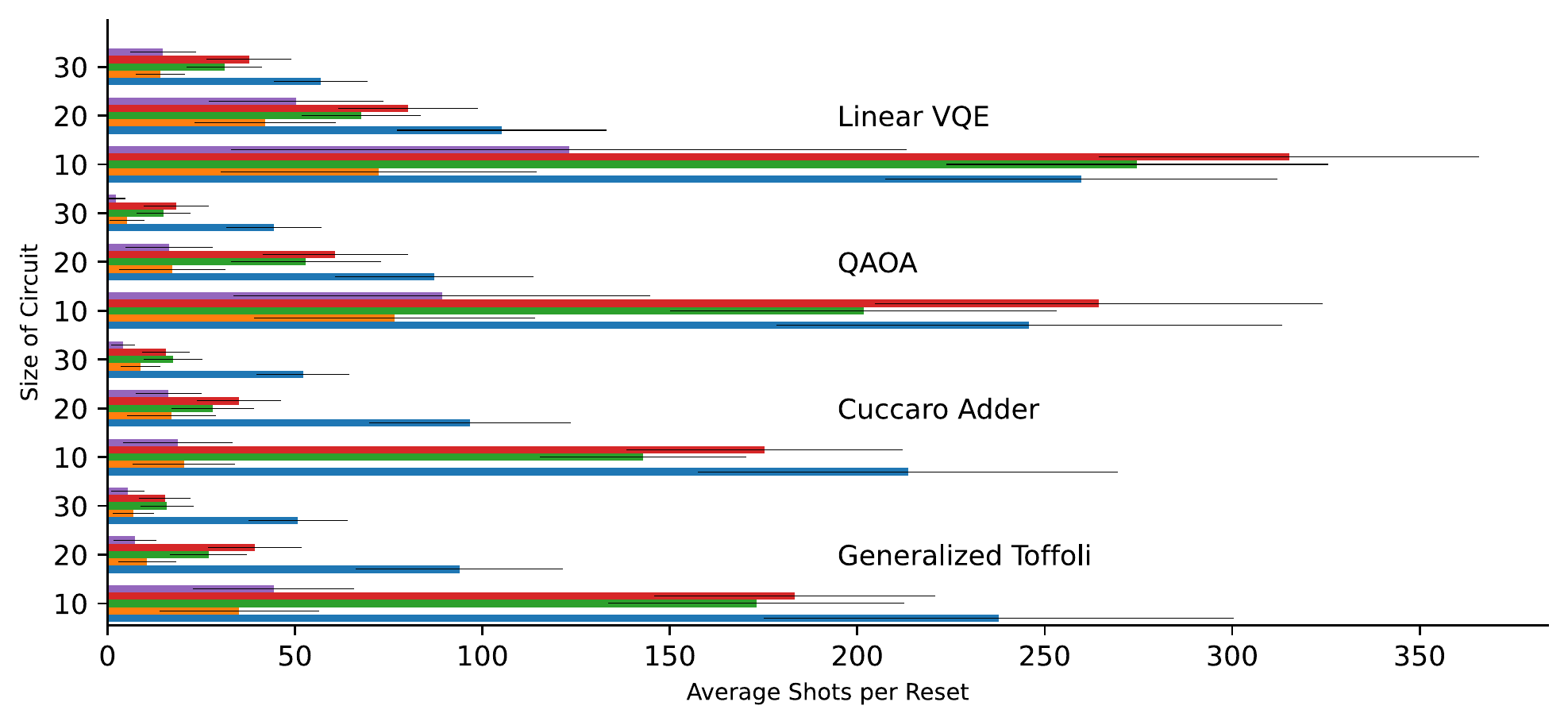%
    }
    \caption{The average number of shots per reload 10, 20 and 30 qubit benchmarks at interaction distance four.  Error bars indicate one standard deviation from the mean.  Each color is a different recovery strategy.  We find that there is significant improvement from our relocation strategies (red and green), greatly exceeding the baseline number of shots, more closely matching recompilation.}
    \label{fig:avg-shots-small}
\vspace{-1.2em}
\end{figure*}

\begin{figure*}
    \centering
    \scalebox{0.75}{
        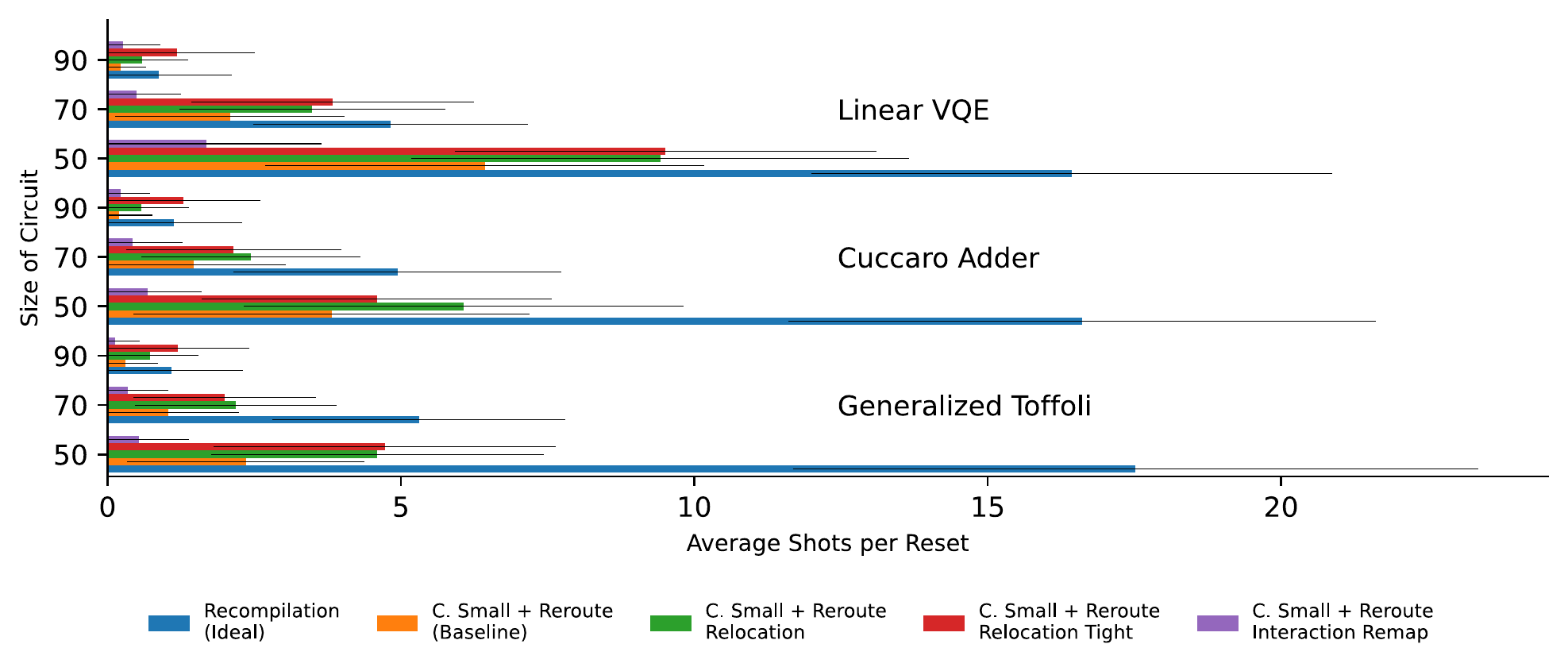%
    }
    \caption{The average number of shots per reload 50, 70 and 90 qubit benchmarks at interaction distance four.  Error bars indicate one standard deviation from the mean.  Each color is a different recovery strategy.  We exclude QAOA since it is not a practical circuit at this size.}
    \label{fig:avg-shots-large}
\vspace{-1.0em}
\end{figure*}

\subsubsection{Focused Use and Relocation}
While shifting along the interaction graph provides benefits for larger circuits, we find that focused use, and relocation to unmapped sections of the architectures can improve probability of success across circuit size, Figures \ref{fig:30-qubit-success-rate} and \ref{fig:90-qubit-success-rate}.  Initially, as atoms are lost, this strategies maintains the same probability of success as our baseline.  This is expected since we are mapping into a specific section of the architecture, we can still generate a similar mapping and routing.  However, after several atoms have been lost, we see an increase in the average probability of success indicating a relocation to a new part of the architecture.  Since there have been fewer atoms lost in this new section there is less adjustment, increasing the probability of success.  Focused use and relocation is able to more closely adhere to the probabilities of success achieved via recompilation, making it a more effective strategy from this perspective.  While we do not achieve strictly better probabilities of success in the 90 qubit circuits, we are able to maintain the same probabilities of success.  We do not see the same benefits since the tiles in this circuit will need to be quite large, at least 9 qubits by 10 qubits.  This requires significant overlap and relocation will not have as great an effect.  Additionally, this strategy does not make use of the interaction graph remapping, and does not benefit from the increased flexibility.

It should also be noted that we do not see a significant difference between the loose and tight bounding boxes.  Both follow the same trend line in each of the benchmarks, indicating that the circuit is mapped and routed similarly in each case.

\subsection{Successful Shots before Reload}

Another metric to consider is how many successful shots can be completed prior to reloading the entire array. Every shot that can be performed without an additional reload reduces the overall time to run a circuit.  We analyze the average successful shots per reload cycle for each mitigation method for each of our benchmarks.

\begin{figure}
    \centering
    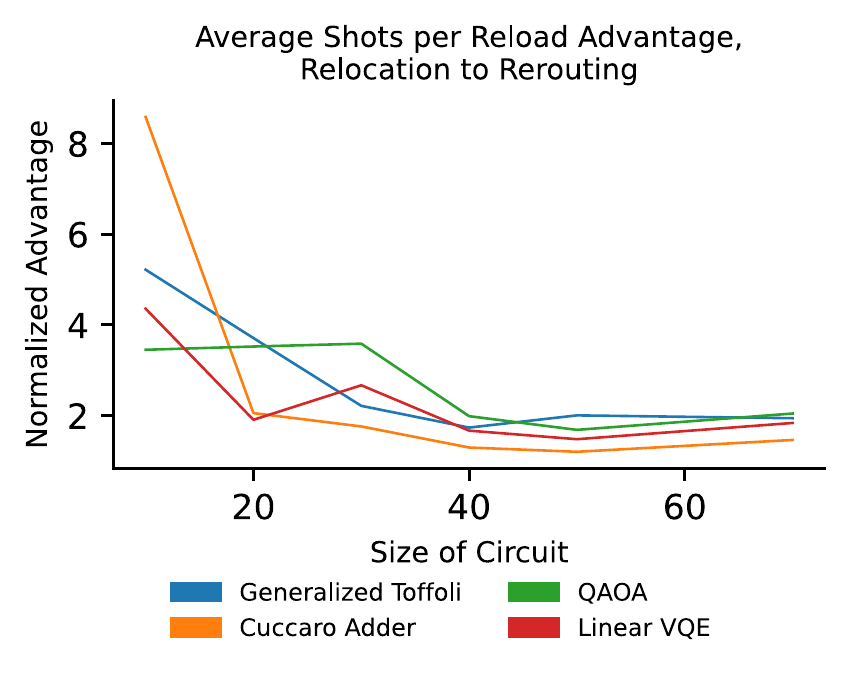%
    \vspace{-1.0em}
    \caption{The normalized advantage of the relocation strategy to rerouting and interaction distance four.  Each color represents a different circuit, and the advantage of relocation decreases at a factor inversely proportional to the size of the circuit.}
    \label{fig:ratio-avg-shots}
\vspace{-1.0em}
\end{figure}

\subsubsection{Remapping Along Interactions}
As seen by the drastically smaller number of average shots per reload, Figure \ref{fig:avg-shots-small}, at smaller circuit sizes remapping by using the interaction graph distance is not as effective as remapping via the architecture graph.  At circuit size 90, Figure \ref{fig:avg-shots-large}, the interaction graph remapping slightly outperforms the array-based remapping. Increased flexibility still provides some recovery for more dense use of the array.

\subsubsection{Focused Use and Relocation}
There is a much more pronounced effect on average shots per reload when utilizing the new relocation strategy over the rerouting strategy for both large and small circuits, Figures \ref{fig:avg-shots-large} and \ref{fig:avg-shots-small}.  We find a much higher rate of average shots per reload for relocation based strategies. Once the array has been divided into several different sections, when the circuit is relocated to a new section it will be mostly free of atom loss. Any atoms lost will be due to computational atoms being remapped into that space, or the rare event of atoms lost to environmental factors and will be much lower. All of which is conducive to much higher probabilities of success and reduces reloads.

As the circuit size increases, we see diminishing returns of the relocation strategy.  As the circuit size increases, the number of distinct sections that can be laid out on the array without overlap, decreases.  We do not have as many opportunities to perform a pseudo-reload as the circuit uses more qubits.  We can even quantify this relationship.  In Figure \ref{fig:ratio-avg-shots} we see the relationship between the increase in the number of shots against the number of times the circuit can be fit onto the architecture.  For each circuit, we follow the same pattern.  The advantage in the number of shots of relocation to rerouting is roughly proportional to the size of the circuit. For a 30 qubit circuit, this is up to 3.5x improvement.  For a 10 qubit circuit, this is up to a 8x improvement in average shots per reload. Empirical results do not match the exact ratio since the bounding box tiles do not fit always fit neatly onto the array. Additionally, these bounding boxes are often larger than the circuit, causing the advantage to be lower.

\subsection{Overhead Time}
\begin{figure}
    \centering
    \scalebox{0.8}{
        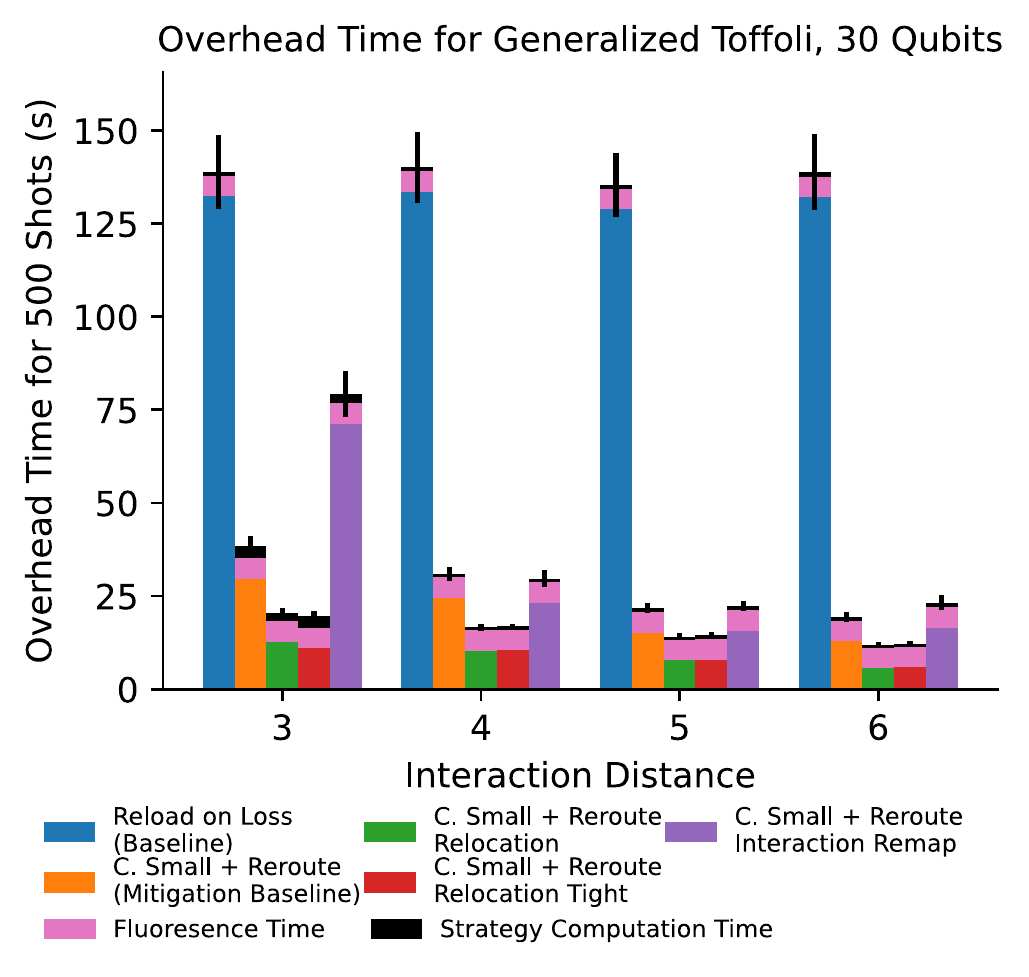%
    }
    \vspace{-1.0em}
    \caption{Overhead times for different recovery strategies at different maximum interaction distances.  The major color in each bar represents time dedicated to reloading.  As follows from the significant advantage seen previously, relocation is a very efficient strategy.}
    \label{fig:overhead-time}
\vspace{-1.0em}
\end{figure}

\begin{figure*}
    \centering
    \scalebox{0.85}{
        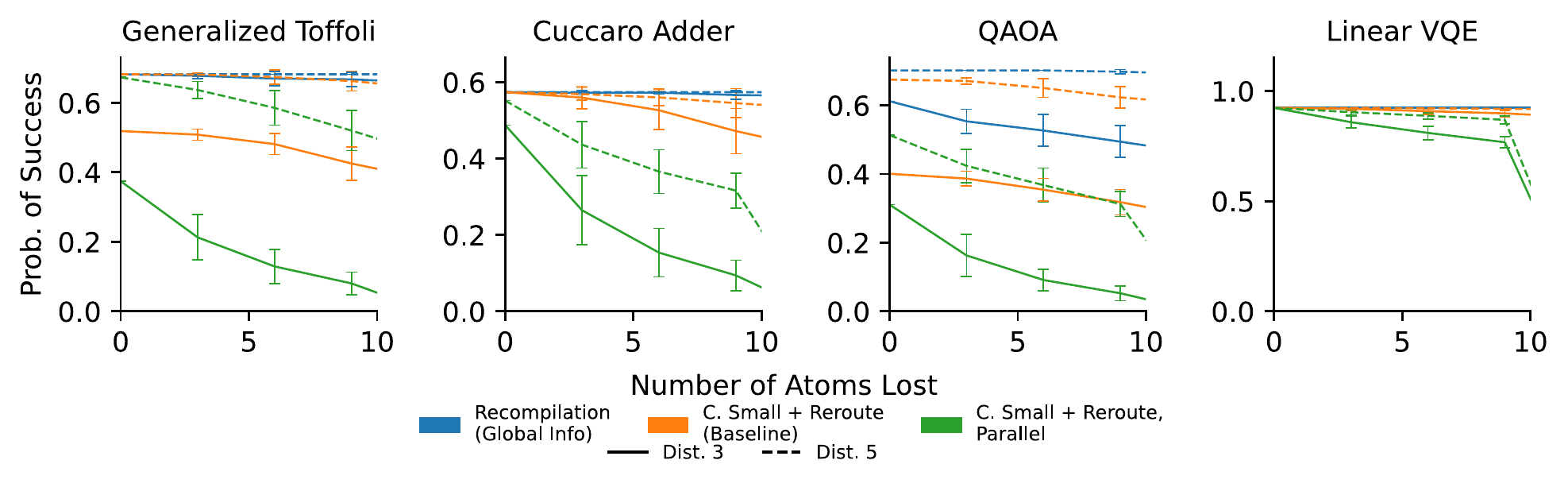%
    }
    \vspace{-1.0em}
    \caption{Decreases in success rate for different mitigation strategies for parallel 30 qubit circuits. Run over 50 trials, each error bar represents one standard deviation from the mean. In general, we see that for higher interaction distances, increased parallelism does not heavily affect the rate of decrease in probability of success rate.}
    \label{fig:prob-success-parallel}
\vspace{-1.5em}
\end{figure*}

\begin{figure*}
    \centering
    \scalebox{0.75}{
        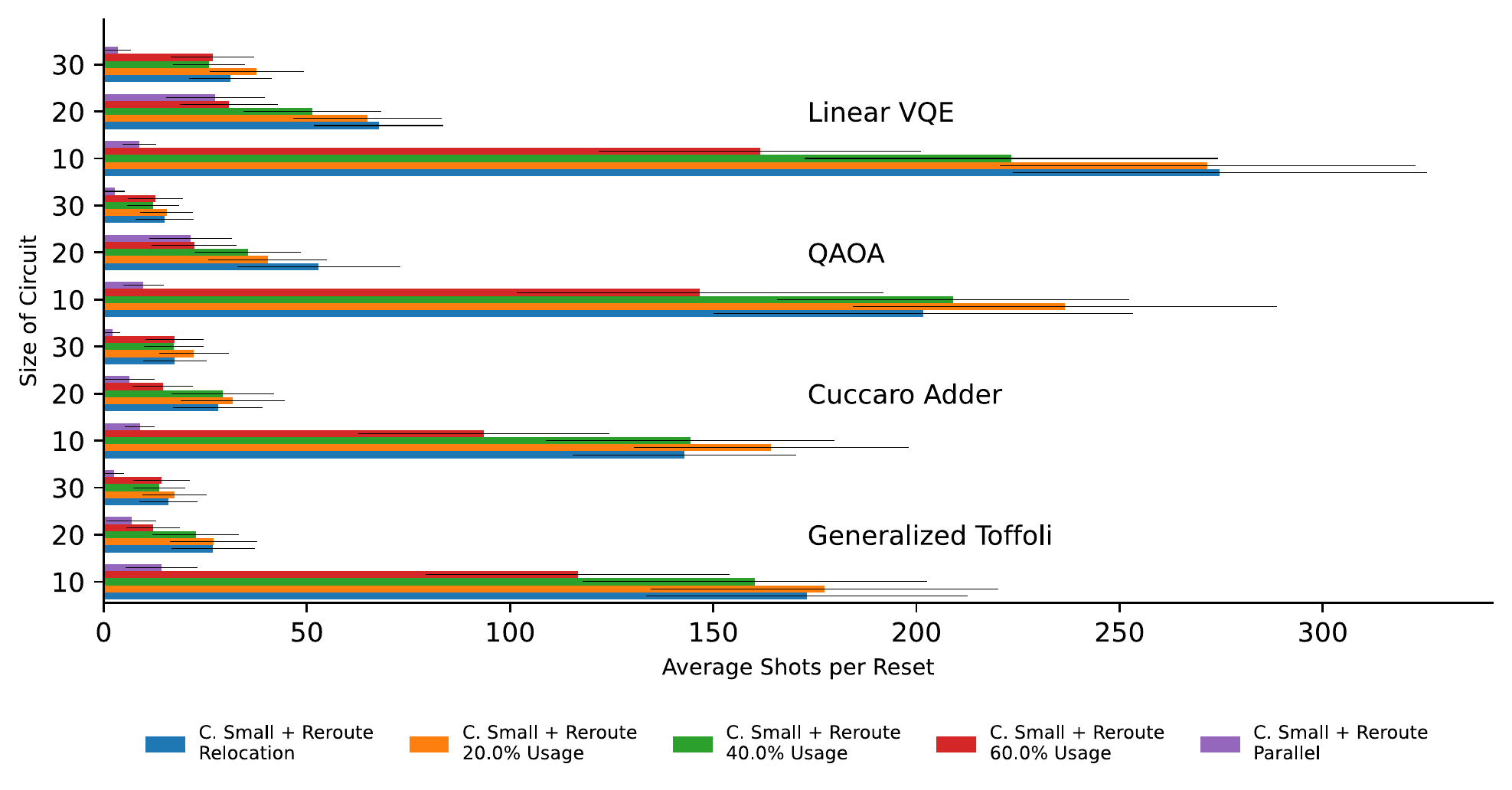%
    }
    \vspace{-1.0em}
    \caption{The average number of shots per reload for 10, 20 and 30 qubit benchmarks at interaction distance four at different levels of parallelism.  Error bars indicate one standard deviation from the mean.  Each color is a level of parallelism.  Parallel indicates that as much the architecture is used as possible, and relocation means that only on instance is run at a time.}
    \label{fig:avg-shots-parallel}
\vspace{-1.0em}
\end{figure*}
While average shots per reload cycle is good indicator for a strategy's potential to recover efficiently, it does not take into account the overhead time to determine the best course of action.  A strategy is only viable if it is faster than reloading the array.  If this cannot be achieved, it is more effective to reload the atoms as it will give us the highest probability of success.  Full recompilation cannot be considered for this reason. Previous work found large gains over reloading via the compiling to a smaller interaction distance and rerouting.  We will be comparing against this strategy to determine effectiveness.

In Figure \ref{fig:overhead-time}, we examine the effects of different mitigation strategies on the overhead time to execute 500 shots for a 30 qubit Generalized Toffoli circuit.  While these strategies are dependent on the length of the circuit, it will scale similarly for each strategy. We see significant reduction in time dedicated to reloading the entire array for our relocation strategies and increases for interaction graph based strategies.  However, the increases in calculating the solutions for relocation do not outweigh the overhead time saved by relocation.  For interaction distance four, our relocation strategy outperforms the basic rerouting strategy by 55\% in reload times, and 45\% overall.

As the maximum interaction distance increases, the margin of the overall time for each strategy decreases.  As the maximum interaction distance increases, we do not need to add communication to recover from an incompatible circuit.  This reduces the number of reloads required for each recovery method. The time dedicated to florescence does not decrease, as each shot, successful or unsuccessful, still requires a fluorescence.

\subsection{Effects of Parallelism}

The benefits of relocation could be improved upon by using multiple non-overlapping sections of the device at the same time.  In doing so, we use as much the architecture as possible. For example, using 90 qubits to run three concurrent 30 qubit circuits.  Since we are using almost all of the qubits, we will use the interaction graph based remapping as it has proven to be more effective in situations where most of the architecture has been filled.  We examine the average probability of success for three concurrent 30 qubit circuits in Figure \ref{fig:prob-success-parallel}. This strategy can only withstand 10 lost qubits at the most since we can only recover as many qubits as those that have been unmapped. Any increased serialization from running circuits in parallel does not substantially impact the decrease in probability of success as compared to previous mitigation strategies.

We also analyze the effects of this strategy on the number of average shots per reload cycle in Figure \ref{fig:avg-shots-parallel}.  By packing as many instances of circuits as possible into the array, there is no space to shift the qubits causing the recovery strategy to fail quickly and requiring a reload.  This frequency will increase the overhead time, indicating that full parallel usage of a neutral atom architecture is not viable at current atom loss rates.

\subsection{Parallelism with Relocation}
Without extra atoms to recover from atom loss, full parallelism is not effective. For smaller circuits, we do not need to fill the entire architecture.  By choosing to only use a portion of the previously defined bounding boxes, we can take advantage of partial parallelism.  By reducing parallelism, we have the opportunity to make use of relocation as well, potentially adding to the overhead time gains we have already found.  In Figure \ref{fig:avg-shots-parallel} we examine multiple levels of parallelism for 10, 20 and 30 qubit circuits.  The percentage of parallelism indicates what percentage of the architecture we are using at a time, 30\% parallelism means that three instances of a 10 qubit circuit are being run at a time, with 3 different areas to relocate these three 10 qubit circuits. We can achieve similar shots per reload to the relocation strategy when running up to four instances of a 10 qubit circuit, and up to two instances of a 20 qubit circuit or 30 qubit circuit.

Since setting up parallelism occurs during the initial compilation phase, this strategy will realize the same gains as relocation in terms of reduction of overhead time due to reload time.  However, by running multiple circuits at the same time, we are executing several shots per fluorescence cycle.  We will have successfully reduced the overhead time from fluorescence by a factor of the number of circuits run at the same time.  This can be seen in Table \ref{tab:parallel-table} for two parallel instances for 10, 20 and 30 qubit circuits.  Put together with the gains from relocation, we see total reductions in overhead time for each of our benchmarks of up to 70\% for a 10 qubit circuit, 60\% for a 20 qubit circuit, and 50\% for a 30 qubit circuit, seen in Table \ref{tab:parallel-table}.


\begin{table}
\begin{center}
\begin{tabular}{l | c | c | c | c}
  Benchmark & Baseline & Relocation & 2 Parallel & \% Fluorescence \\
   & Time (s) & Time (s) & Time (s) & Decrease \\
\hline
  Cuccaro-10 & 10.91 & 4.67 & 3.13 & 51.22 \\
  Cuccaro-20 & 13.82 & 9.58 & 8.24 & 54.02 \\
  Cuccaro-30 & 26.19 & 17.05 & 14.80 & 40.71 \\
  CNU-10 & 7.98 & 4.59 & 2.82 & 50.60 \\
  CNU-20 & 24.09 & 8.68 & 10.50 & 53.12 \\
  CNU-30 & 30.61 & 16.58 & 15.15 & 35.80 \\
  QAOA-10 & 5.83 & 4.41 & 2.66 & 51.05 \\
  QAOA-20 & 14.83 & 7.54 & 7.49 & 52.70 \\
  QAOA-30 & 42.67 & 16.43 & 18.68 & 35.48 \\
  Linear VQE-10 & 6.26 & 4.16 & 2.42 & 50.83 \\
  Linear VQE-20 & 8.77 & 6.61 & 5.01 & 53.75 \\
  Linear VQE-30 & 15.93 & 9.79 & 9.56 & 36.12 \\
\end{tabular}
\caption{Full Runtimes with Two Instances of Parallelism}
\label{tab:parallel-table}
\vspace{-1.5em}
\end{center}
\end{table}
\section{Discussion and Conclusion}
Neutral atom systems are a potential contender for scalable quantum computing.  However, the underlying technology is not without its problems.  Atom loss is a challenge towards reaching scalability. As architecture and viable circuit sizes increase, atom loss prevents fast execution of repeated successful shots.  Through software techniques this overhead can be mitigated.

Initial techniques for reducing overhead focus on adjusting the initial mapping and routing of a circuit on a device.  This is effective when there is enough open space on an architecture.  As resources become more constrained, we require more flexible techniques.  By exploiting the unique feature of long distance interactions, we can make use of a small number of atoms to improve how many successful shots can be achieved before reloading the circuit via the interaction graph.  This flexibility gives an escape valve for more edge case circuits.

However, we do not only focus on adjusting the initial mapping and routing.  By changing the initial mapping and routing while focusing on effectively using all of an architecture and more actively avoiding lost atoms,  we can further reduce overhead time.  Through the creation of tiled bounding boxes, dividing the array into several small architectures, and fully exhausting each of these sections before moving into a new section without atom loss overhead time due to reloading the array is reduced by a factor of the number of times the circuit fits on the architecture.  This can all be done while keeping the probability of success high and without significant increases in the time to adapt the circuit to the lost atoms.

Finally, we also use these bounding box sections to exploit parallelism.  While using every available atom on a device is not effective due to space constraints, using less than 100\% of the architecture combined with relocation achieves the same number of shots per reload while also reducing the number of runs of the array required.  This reduces the overhead time dedicated to florescence, further improving our ability to quickly run repeated shots on the neutral atom device.

Together, these techniques build a more complete set of strategies to handle different program sizes and circuit structures as neutral atoms are further developed.  One strategy is not usually enough to handle every case and this ensemble of techniques use the unique aspects of the architecture to help realize its full potential. 

\bibliography{bibliography}

\end{document}